\documentclass[twocolumn,showpacs,preprintnumbers,superscriptaddress,amsmath,amssymb,a4paper]{revtex4}

\usepackage{amsmath,amssymb,bbm,epsfig}
\usepackage{rotating}

\newcommand{\be}{\begin{equation}}
\newcommand{\ee}{\end{equation}}
\newcommand{\ba}{\begin{array}}
\newcommand{\ea}{\end{array}}
\newcommand{\bqa}{\begin{eqnarray}}
\newcommand{\eqa}{\end{eqnarray}}

\newcommand{\um}{\mathbbm{1}}

\newcommand{\bra}[1]{\ensuremath{\langle #1 |}}
\newcommand{\ket}[1]{\ensuremath{| #1 \rangle}}

\newcommand{\newket}[1]{\ensuremath{\left|\ba{c} #1 \ea\right\rangle}}

\begin{document}

\title{Many-Body Entanglement: Permutations and Equivalence classes}

%\author{Florian Mintert, Benno Salwey, Andreas Buchleitner}
%\iffalse
\author{Florian Mintert}
\altaffiliation{Freiburg Institute for Advanced Studies, Albert-Ludwigs-Universit\"at Freiburg, Albertstr. 19, D-79104 Freiburg, Germany}
%\affiliation{Freiburg Institute for Advanced Studies, Albert-Ludwigs-Universit\"at Freiburg, Albertstr. 19, D-79104 Freiburg, Germany}
\affiliation{Physikalisches Institut, Albert-Ludwigs-Universit\"at Freiburg, Hermann-Herder-Strasse 3, D-79104 Freiburg, Germany}
\author{Benno Salwey}
\altaffiliation{D\'epartement IRO, Universit\'e de Montr\'eal C.P. 6128, succursale centre-ville Montr\'eal (Qu\'ebec), Canada H3C 3J7}
\affiliation{Physikalisches Institut, Albert-Ludwigs-Universit\"at Freiburg, Hermann-Herder-Strasse 3, D-79104 Freiburg, Germany}
\author{Andreas Buchleitner}
\affiliation{Physikalisches Institut, Albert-Ludwigs-Universit\"at Freiburg, Hermann-Herder-Strasse 3, D-79104 Freiburg, Germany}
%\fi

\date{\today}

\begin{abstract}
 With an easily applicable criterion based on permutation symmetries of (identically prepared) replicas of quantum states we identify distinct entanglement classes in high-dimensional multi-partite systems.
The different symmetry properties of inequivalent states provide a rather intuitive picture of the otherwise very abstract classification of many-body entangled states.
\end{abstract}

\pacs{
03.65.Ud,	%Entanglement and quantum nonlocality
03.67.Mn,	%Entanglement measures, witnesses, and other characterizations
03.65.Fd} % Algebraic methods}
%02.10.De	Algebraic structures and number theory,
%02.10.Ud	Linear algebra}

\maketitle

%{\bf map on/to/onto ?}

%\watermark

Many-body quantum coherence is responsible for many astonishing properties of composite quantum systems, what concerns their spectral, {\it e.g.}, specifically ground state \cite{sachdev} properties but also their quantum dynamics: 
think of superconductivity where coherently composed quantum systems function as carriers that guarantee lossless charge transport \cite{PhysRev.106.162}, or of enhanced exitation transport in photosynthetic processes, which is under debate to be of quantum origin \cite{HohjaiLee06082007}. 
The inherent many-body character of such features makes them difficult to understand in terms of single particle observables which are the categories of our classically trained intuition. Consequently, we need a better understanding of the nature of many-body quantum coherence, {\it i.e.} of multi-partite entanglement.

So far, the structure of entangled states is still widely unexplored. Only for pure states of bipartite systems do we have an exhaustive understanding based on the Schmidt decomposition  \cite{schmidt_deco}. But in multipartite systems, where there is no comparably useful tool, we have only very partial knowledge. 
We know that there exist two inequivalent classes of entangled states in three qubit systems  \cite{PhysRevA.62.062314}, the GHZ and W states. Four qubit systems reveal even a number of distinct families, i.e. continuously parametrized sets of classes, of entangled states \cite{PhysRevA.65.052112}.
Further efforts \cite{PhysRevA.69.012101,PhysRevA.73.032314,arXiv:0911.1803,osterloh,arXiv:1001.0078}, still restrict to rather small and low-dimensional systems or to states with specific symmetry properties \cite{Photon_Classification}, which hitherto leaves us with a rather intransparent picture of the general structure of multipartite entanglement.

The defining property of entanglement is that it can not be created
by local manipulation.
Formally, this is expressed in terms of a Kraus operator $K$ \cite{kraus,chuang00}:
if there is an operator $K$ such that $\ket{\Phi}=K\ket{\Psi}$,
then there is a generalized measurement, also referred to as POVM \cite{chuang00}, in which one of the outcomes is associated with the transition of the state $\ket{\Psi}$ to $\ket{\Phi}$.
If the Kraus operator acting on an $N$-partite system is a simple direct product of local Krauss operators $F_i$,
{\it i.e.} if $K=F_1\otimes\hdots\otimes F_N$,
then the measurement can be implemented in terms of local measurements on the $N$ subsystems.

In bipartite systems a state $\ket{\Phi}$ can be created from a state $\ket{\Psi}$ by local manipulation
if and only if the Schmidt rank, {\it i.e.} the number of finite Schmidt coefficients of $\ket{\Phi}$, does not exceed that of $\ket{\Psi}$.
Thus, starting out from a state with maximal Schmidt rank in a given finite dimensional bipartite system,
{\em any} other state can be obtained through local manipulation only.
This is fundamentally different in multipartite systems where there is in general no such so-called maximally entangled state, but there are distinct classes or families of multipartite states.
Two states $\ket{\Phi}$ and $\ket{\Psi}$ define two different equivalence classes if there are no local Kraus operators $F_i$ ($i=1,...,N$) and $G_i$ ($i=1,...,N$)
such that $\ket{\Phi}=F_1\otimes\hdots\otimes F_N\ket{\Psi}$ and $\ket{\Psi}=G_1\otimes\hdots\otimes G_N\ket{\Phi}$.
In tripartite qubit systems there are the two classes of $\ket{W}=(\ket{211}+\ket{121}+\ket{112})/\sqrt{3}$ and $\ket{GHZ}=(\ket{111}+\ket{222})/\sqrt{2}$ states \cite{PhysRevA.62.062314},
and in fourpartite qubit systems there are even continuous families of inequivalently entangled states \cite{PhysRevA.65.052112}.
Only if the dimension of one of the subsystems is as least as large as the product of the dimensions of all other subsystems there is a maximally entangled state in the above sense \cite{arXiv:0911.0879}.

Our task here is to describe how permutation symmetries of multiple quantum states can give us an intuitive understanding of the otherwise very abstract classification of SLOCC equivalence classes.
For this purpose we will first discuss in detail how known classifications can be revealed through such symmetries, and then identify new classes.
For the latter, permutation symmetries first of all allow us to intuitively identify states with inequivalent properties,
and, once identified, they also permit the rigorous proof of their inequivalence. 

%------------------------------------------------------------------------------------------------------------------------------

\section{Replicas of quantum systems}
%%%%%    Benno 24.2.2011 : anderer Vorschlag weil Permutations evtl. noch genereller als irreduz. Darstellungen
%{\bf We will rephrase the the criterion  \cite{PhysRevA.77.052304}, which we will exploit here to identify structure of many-body states, in general terms of permutation symmetries of multiple replica of a quantum state.}
 The criterion \cite{PhysRevA.77.052304} that we exploit here to identify structure of many-body entangled states is based on permutation symmetries of multiple replicas of a quantum state.
`Multiple' in the present context means that rather than considering a quantum state $\ket{\Phi}$ of an $N$-partite quantum system,
we consider $M$ replicas of this state vector, {\it i.e.} the $M$-fold tensor product of the state vector $\ket{\Phi}$ with itself.
Such permutation symmetries must not be mistaken with symmetries under permutations of subsystems which are frequently considered \cite{PhysRevA.77.012104,PhysRevLett.103.020504,PhysRevLett.102.170503,arXiv:1001.0343}.
In order to avoid such confusion between the different subsystems and different replicas of a quantum state,
we will adopt the convention of \cite{PhysRevA.75.022328} to write the different replicas of a quantum state in different rows.
All terms associated with a specific subsystem will the be within a column.
That is, if we consider a bipartite quantum state
$\ket{\Psi}=\sum_{ij}\Psi_{ij}\ket{i}\otimes\ket{j}$,
then two replicas read
$\ket{\Psi}^{\otimes 2}=\sum_{i_1i_2j_1j_2}\Psi_{i_1j_1}\Psi_{i_2j_2}\ket{i_1}\otimes\ket{j_1}\otimes\ket{i_2}\otimes\ket{j_2}$,
and we will denote the four-component state vector as
\be\ba{cccccc}
\ket{i_1}&\otimes&\ket{j_1}&\hspace{.5cm}&\in{\cal H}_1\otimes{\cal H}_2\\
\otimes&&\otimes\\
\ket{i_2}&\otimes&\ket{j_2}&\hspace{.5cm}&\in{\cal H}_1\otimes{\cal H}_2\vspace{.5cm}\\
\begin{sideways}$\ni$\end{sideways}&&\begin{sideways}$\ni$\end{sideways}\\
{\cal H}_1&&{\cal H}_2\\
\otimes&&\otimes\\
{\cal H}_1&&{\cal H}_2&,
\ea
\label{eq:twofoldbipart}
\ee
or, in short hand notation, as
\be
\left|\ba{c}i_1\\i_2\\\ea\right\rangle\otimes\left|\ba{c}j_1\\j_2\\\ea\right\rangle\ .
\ee

%------------------------------------------------------------------------------------------------------------------------------

\subsection{Permutations}

As we will discuss in the remainder of this article we can use permutations on several replicas of a subsystem to distinguish different classes of entangled states.
In the case of two replicas, the (unique) permutation operator $\Pi$ associated with one subsystem is defined via
\be
\Pi\left|\ba{c}k_1\\k_2\\\ea\right\rangle=\left|\ba{c}k_2\\k_1\\\ea\right\rangle\ ,
\label{eq:Pi}
\ee
and, in the case of  $M$ replicas
there are $M!$ inequivalent permutations \cite{PhysRevA.58.1833} per subsystem that are defined similarly to Eq.~\eqref{eq:Pi}.

 In the case of \eqref{eq:twofoldbipart} there are two distinct permutations possible: one that interchanges $\ket{i_1}$ and $\ket{i_2}$ on subsystems one and another one that interchanges $\ket{j_1}$ and $\ket{j_2}$ on the subsystems two.
Any separable state $\ket{\Psi_1}=\ket{\varphi_1}\otimes\ket{\varphi_2}$ is invariant under either of these permutations:
\be\ba{rcl}
\ket{\Psi_1}^{\otimes 2}=\newket{\varphi_1\\\varphi_1}\otimes\newket{\varphi_2\\\varphi_2}&=&\Pi\newket{\varphi_1\\\varphi_1}\otimes\newket{\varphi_2\\\varphi_2}\vspace{.2cm}\\
&=&\newket{\varphi_1\\\varphi_1}\otimes\Pi\newket{\varphi_2\\\varphi_2}\ .
\ea\ee
This is different for an entangled state, that we can write in its Schmidt decomposition as
$\ket{\Psi_2}=\sqrt{\lambda_1}\ket{11}+\sqrt{\lambda_2}\ket{22}$.
Its duplicate version
reads
\be\ba{r}
\ket{\Psi_2}^{\otimes 2}=\lambda_1\left|\ba{c}1\\1\\\ea\right\rangle\otimes\left|\ba{c}1\\1\\\ea\right\rangle+\lambda_2\left|\ba{c}2\\2\\\ea\right\rangle\otimes\left|\ba{c}2\\2\\\ea\right\rangle+\vspace{.2cm}\\
\sqrt{\lambda_1\lambda_2}\left(\left|\ba{c}1\\2\\\ea\right\rangle\otimes\left|\ba{c}1\\2\\\ea\right\rangle+\left|\ba{c}2\\1\\\ea\right\rangle\otimes\left|\ba{c}2\\1\\\ea\right\rangle\right)\ .
\ea\ee
This object is actually altered by a permutation on the first subsystem:
\be\ba{r}
\ket{\Psi_2}^{\otimes 2}\neq\lambda_1\left|\ba{c}1\\1\\\ea\right\rangle\otimes\left|\ba{c}1\\1\\\ea\right\rangle+\lambda_2\left|\ba{c}2\\2\\\ea\right\rangle\otimes\left|\ba{c}2\\2\\\ea\right\rangle+\vspace{.2cm}\\
\sqrt{\lambda_1\lambda_2}\left(\left|\ba{c}2\\1\\\ea\right\rangle\otimes\left|\ba{c}1\\2\\\ea\right\rangle+\left|\ba{c}1\\2\\\ea\right\rangle\otimes\left|\ba{c}2\\1\\\ea\right\rangle\right)\ ,
\ea\ee
and similarly for a permutation on the second subsystem.
Consequently, the projector $P_-=(\um-\Pi)/2$ onto the antisymmetric component of ${\cal H}_1\otimes{\cal H}_1$ ( or analogously of ${\cal H}_2\otimes{\cal H}_2$ ) can now be used to probe the invariance of any given duplicate state
$\ket{\Psi}\otimes\ket{\Psi}$ under $\Pi$:
any separable state is invariant under $\Pi$ and is thus mapped on the zero-vector
\be
(P_-\otimes\um)\ket{\Psi_1}^{\otimes 2}=0\ ,
\ee
whereas any entangled state remains finite:
\be\ba{l}
(P_-\otimes\um)\ket{\Psi_2}^{\otimes 2}=
\frac{\sqrt{\lambda_1\lambda_2}}{2}\left(
\newket{1\\2}\otimes\newket{1\\2}-\right.\vspace{.2cm}\\
\left.-\newket{2\\1}\otimes\newket{1\\2}+
\newket{2\\1}\otimes\newket{2\\1}-\newket{1\\2}\otimes\newket{2\\1}\right)\ .
\label{eq:septoent}
\ea\ee

Later-on, we will see that similar distinctions also exist for different classes of \emph{multipartite} entangled states.
However, before doing so, let us formulate our criterion to identify inequivalent entanglement classes.

%------------------------------------------------------------------------------------------------------------------------------

\subsection{Criterion for the classification of many-body entanglement}

The central property of any permutation operator $\Pi$ acting on $M$ replicas of a single-particle Hilbert space is that it commutes with the $M$-fold tensor product of {\em any} operator $F$:
\be
[\Pi,F^{\otimes M}]=0\ .
\ee
Permutation operators are ideally suited to distinguish different classes of entangled states, since this relation naturally generalizes to $M$ replicas of an $N$-partite system, if we introduce introduce the operator
\be
A=\sum_{{i_1\hdots i_N}\atop{j_1\hdots j_N}}\eta_{i_1\hdots i_N}^{j_1\hdots j_N}\ \Pi_{i_1}^{(j_1)}\otimes\hdots\otimes\Pi_{i_N}^{(j_N)}.
\label{eq:A}
\ee
$A$ is composed of permutations $\Pi_i^{(j)}$, where `$j$' denotes the subsystem that the permutation is associated with,
`$i$' labels different permutations,
and $\eta_{i_1\hdots i_N}^{j_1\hdots j_N}$ are complex numers.
Any such operator $A$ commutes with with $M$ replicas of a local Kraus operator
\be
[A,(F_1\otimes\hdots\otimes F_N)^{\otimes M}]=0\ .
\label{eq:commute}
\ee
With this at hand, we can rephrase the criterion for the classification of many-body entanglement \cite{PhysRevA.77.052304} that we will exploit in the following:
\begin{quote}
\emph{
If an operator $A$ as specified in Eq.~\eqref{eq:A} exists such that
\be
A\ket{\Psi}^{\otimes M}=0\ ,\ \mbox{and}\ A\ket{\Phi}^{\otimes M}\neq 0\ ,
\label{eq:cond}
\ee
then the state $\ket{\Phi}$ cannot be obtained from $\ket{\Psi}$ through local manipulation.
}
\end{quote}

The proof of this assertion follows by contradiction:
\bqa
0\neq A\ket{\Phi}^{\otimes M}&=&A\bigl((F_1\otimes\hdots\ F_N)\ket{\Psi}\bigr)^{\otimes M}\\
&=&(F_1\otimes\hdots\ F_N)^{\otimes M} A \ket{\Psi}^{\otimes M}=0\nonumber\ ,
\eqa
as a direct consequence of Eqs.~\eqref{eq:commute} and \eqref{eq:cond}.

There are cases in which a state $\ket{\Phi}$ can be created through local action on a state $\ket{\Psi}$, but where the inverse is not possible.
In these cases it is meaningful to assert that $\ket{\Psi}$ carries entanglement of superior type than the the state $\ket{\Phi}$.
However, as it is the case for example with the $GHZ$ and the $W$ state, there are also different classes of entangled states and neither can one state of the former class be obtained through local manipulation of a state of the latter, nor is the reverse possible.
To show such distinctness
% an inequivalence-relation {\bf terminology}
for two states $\ket{\Phi}$ and $\ket{\Psi}$ with the help of Eq.~\eqref{eq:cond},
two operators $A_1$ and $A_2$ are required with
\bqa
A_1\ket{\Psi}=0 & \mbox{and} & A_1\ket{\Phi}\neq 0\ ,\ \mbox{but}\\
A_2\ket{\Psi}\neq 0 & \mbox{and} & A_2\ket{\Phi}=0\ .
\eqa
However, if the ranks of all single-subsystem reduced density matrices of both states $\ket{\Phi}$ and $\ket{\Psi}$ coincide
for any subsystem, Eq.~\eqref{eq:cond} is already sufficient for distinctiveness:
in this case all local Kraus operators can be assumed to be invertible, without loss of generality \cite{PhysRevA.62.062314},
so that the non-existence of local operators $F_i$ ($i=1,...,n$) with $\ket{\Phi}=F_1\otimes\hdots\otimes F_n\ket{\Psi}$
implies the non-existence  of local operators $G_i$ ($i=1,...,n$) with $\ket{\Psi}=G_1\otimes\hdots\otimes G_n\ket{\Phi}$.

%------------------------------------------------------------------------------------------------------------------------------

\section{Classification of entangled many-body states}

Before using the above formalism to identify new structures within the set of multi-partite states, we will first discuss how earlier results can be reproduced, in order to gain some insight in this technique.
Then we will demonstrate with a few examples how the employed permutation symmetries can be used to identify hitherto unknown inequivalences even in the context of systems of unbounded size (in dimension and number of subsystems). The latter is due to the independence of Eq.~\eqref{eq:cond} of the systems size and becomes possible once an intuition for the action of the operators in Eq.~\eqref{eq:A} on entangled states of certain symmetries is developed.

\subsection{The bipartite case revisited}
\label{subsec:bipartite}

With Eq.~\eqref{eq:septoent} we have verified that no entangled bi-qubit states $\ket{\Psi_2}$ can be created by local action on a separable state $\ket{\Psi_1}$.
To show that in a system of higher dimensional constituents a state $\ket{\Psi_3}=\sqrt{\lambda_1}\ket{11}+\sqrt{\lambda_2}\ket{22}+\sqrt{\lambda_3}\ket{33}$ with three finite Schmidt coefficients can not be obtained from a state with only two finite Schmidt coefficient, {\it i.e.} a state with a Schmidt decomposition as $\ket{\Psi_2}$,
we have to consider three replicas of this quantum state.
In this framework we can define the projector ${\cal A}_3$
onto the completely antisymmetric (fermionic) subspace of a triplicate single-particle Hilbert space,
%${\cal H}_1\otimes{\cal H}_1\otimes{\cal H}_1$, 
{\it i.e.}
\be
{\cal A}_3=\frac{1}{3!}\sum_{i=1}^{3!}\mbox{sgn}(\Pi_{i})\ \Pi_{i}\ .
\label{eq:defantisymmetrizer3}
\ee
Here `sgn' denotes the signature,
{\it i.e.} the function that yields sgn$(\Pi_e)=1$ for any permutation $\Pi_e$ that can be decomposed into an even number of pairwise permutations (transpositions),
and sgn$(\Pi_o)=-1$ for any permutation $\Pi_o$ that can be decomposed into an odd number of pairwise permutations.
%%%%%    Benno 24.2.2011 : ein 'd' an decompose gehaengt
Let us inspect the action of ${\cal A}_3$ on a three-replica state-vector.
If the components of the three replicas are pairwise different, then the corresponding state is mapped on the completely anti-symmetric state
\be
\ket{{\cal A}}=\frac{1}{\sqrt{6}}\left(\left|\ba{c}1\\2\\3\\\ea\right\rangle-\left|\ba{c}1\\3\\2\\\ea\right\rangle-\left|\ba{c}2\\1\\3\\\ea\right\rangle+\left|\ba{c}2\\3\\1\\\ea\right\rangle+\left|\ba{c}3\\1\\2\\\ea\right\rangle-\left|\ba{c}3\\2\\1\\\ea\right\rangle\right)
\ee
by ${\cal A}_3$, {\it i.e.}
\be
{\cal A}_3\left|\ba{c}i\\j\\k\\\ea\right\rangle=\frac{1}{\sqrt{6}}\mbox{sgn}(ijk)\ket{{\cal A}}\ ,
\label{eq:signature}
\ee
where the signature $\mbox{sgn}$ of a string of numbers is defined similarly to the signature of a permutation above,
{\it i.e.} $\mbox{sgn}(123)=\mbox{sgn}(231)=\mbox{sgn}(312)=1$ and $\mbox{sgn}(213)=\mbox{sgn}(321)=\mbox{sgn}(132)=-1$.
If $i$, $j$ and $k$ are not pairwise different, then the state-vector is mapped on the zero-vector by ${\cal A}_3$.

Since three replicas of a state $\ket{\Psi_2}$ contain no term with three pairwise different contributions per subsystem,
the three replicas $\ket{\Psi_2}^{\otimes 3}$ are necessarily mapped on the zero-vector by the application of ${\cal A}_3$ on either subsystem. 
But, the three-level entangled state $\ket{\Psi_3}$ behaves differently:
three replicas contain the component
\be
\sqrt{\lambda_i\lambda_j\lambda_k}\newket{i\\j\\k}\otimes\newket{i\\j\\k}\ ,
\ee
with $i\neq j\neq k\neq i$, that survive the application of ${\cal A}_3$, such that
\be
{\cal A}_3\otimes\um\ket{\Psi_3}^{\otimes 3}=\sqrt{\lambda_1\lambda_2\lambda_3}\sum_{ijk}\frac{\mbox{sgn}(ijk)}{\sqrt{6}}\ket{{\cal A}}\otimes\newket{i\\j\\k\\}\neq 0\ .
\label{eq:truethreelevelent}
\ee
Together with Eq.~\eqref{eq:cond} this verifies that $\ket{\Psi_3}$ cannot be obtained from $\ket{\Psi_2}$ through local manipulation.
With exactly the same argument, one can also show that a state $\ket{\Psi_N}$ with $N$ finite Schmidt coefficients can not be obtained from states with fewer finite Schmidt coefficients,
simply invoking the projector
\be
{\cal A}_N=\frac{1}{N!}\sum_{i=1}^{N!}\mbox{sgn}(\Pi_{i})\ \Pi_{i}\ ,
\label{eq:AN}
\ee
onto the completely antisymmetric part of ${\cal H}_1^{\otimes N}$.
Doing so, one recovers the full classification of bipartite entanglement in terms of the Schmidt rank \cite{QInfCompNielsenVidal}.

%------------------------------------------------------------------------------------------------------------------------------

\subsection{$GHZ$ states vs. $W$ states}

The inequivalence of $GHZ$ and $W$ states can be shown in a similar manner.
Rephrasing the three-qubit tangle $\tau$ \cite{PhysRevA.61.052306} in terms of permutation operators, one finds that
$\tau(\Psi)=16\sqrt{\bra{\Psi}^{\otimes 4}A_\tau\ket{\Psi}^{\otimes 4}}$ with
\be
A_\tau=\underbrace{P_-^{(12)}\otimes P_-^{(34)}}_{{\cal H}_1^{\otimes 4}}\otimes\underbrace{P_-^{(12)}\otimes P_-^{(34)}}_{{\cal H}_2^{\otimes 4}}\otimes\underbrace{P_-^{(13)}\otimes P_-^{(24)}}_{{\cal H}_3^{\otimes 4}}\ ,
\label{eq:Atau}
\ee
where two consecutive projectors act on the four replicas of a subsystem as depicted above.
The fact that $A_\tau\ket{GHZ}^{\otimes 4}$ is finite is implied by the finite tangle of the $GHZ$ state,
but, one also readily convinces oneself that
\be
A_\tau\ket{GHZ}^{\otimes 4}=\frac{1}{4}\ket{\xi_\tau}\ ,
\label{eq:AGHZ}
\ee
where $\ket{\xi_\tau}$ is the unique eigenvector of $A_\tau$ associated with the unit eigenvalue.
For this purpose, let us take a look at those terms in $\ket{GHZ}^{\otimes 4}$ that are not mapped on the zero vector by $A_\tau$.
The projection $P_-^{(12)}\otimes P_-^{(34)}$ on the first subsystem replica spaces requires the terms associated with the first and second replica as well as those associated with the third and fourth replica to be different.
In turn, $P_-^{(13)}\otimes P_-^{(24)}$ on the third subsystem enforces terms associated with the first (second) and third (fourth) replica to be different. 
The only terms in $\ket{GHZ}^{\otimes 4}$ that satisfy these requirements simultaneously are of the form
\be
\ket{\kappa_{ij}}=\newket{i\\j\\j\\i}\otimes\newket{i\\j\\j\\i}\otimes\newket{i\\j\\j\\i}\ ,
\label{eq:termsGHZ}
\ee
with $i\neq j$ ({\it i.e.} all other terms vanish one by one under the application of $A_\tau$). 
What remains to be verified is that the term with $i=1,j=2$ does not cancel the term with $i=2,j=1$, {\it i.e.}
\be
\sum_{ij}A_\tau\ket{\kappa_{ij}}\neq 0\ .
\ee
Both state-vectors $\newket{1\\2}$ and $\newket{2\\1}$ are mapped on the singlet state by $P_-$,
but with different phases:
\be
P_-\newket{1\\2}=-P_-\newket{2\\1}=\frac{1}{2}\left(\newket{1\\2}-\newket{2\\1}\right)=\frac{1}{\sqrt{2}}\ket{\chi}\ .
\ee
In Eq.~\eqref{eq:termsGHZ}, both of the terms $\newket{i\\j}$ and $\newket{j\\i}$ appear threefold:
$\newket{i\\j}$ appears in the first and second replica component of the first and second subsystem,
and in the first and third replica component of the third subsystem;
$\newket{j\\i}$ appears in the third and fourth replica component of the first and second subsystem,
and in the second and fourth replica component of the third subsystem.
That is, in both of the two cases  $i=1,j=2$ and $i=2,j=1$, the application of $A_\tau$, which is composed of $6$ terms $P_-$, yields a sixfold tensor product of $\ket{\chi}$ with three times the negative prefactor
$-1/\sqrt{2}$, and three times the positive prefactor $1/\sqrt{2}$.
That is, in both of these cases, the same resulting prefactor $-1/8$ is obtained, so that these two terms add up constructively, {\it i.e.} do \emph{not} cancel each other.

As we discuss in the following, the $\ket{W}$ state behaves differently, {\it i.e.}
$A_\tau\ket{W}^{\otimes 4}=0$.
To verify this, let us introduce the short hand notations $\ket{\omega_1}=\ket{211}$, $\ket{\omega_2}=\ket{121}$ and $\ket{\omega_3}=\ket{112}$.
Now, the quadruplication of $\ket{W}$ is comprised of $3^4$ terms, but we can easily verify that each of these terms is mapped on the zero-vector by $A_\tau$.
The projector $P_-^{(12)}\otimes P_-^{(34)}$ acting on the replicas of the first subsystem, maps all those terms onto the zero-vector that do not
contain $\ket{\omega_1}$ exactly once in the first two replicas and exactly once in both the third and fourth replicas.
Similarly, the projector $P_-^{(12)}\otimes P_-^{(34)}$, acting on the second subsystem, maps all those terms onto the zero-vector that do not
contain $\ket{\omega_2}$ exactly once in the first two replica and exactly once in both the third and fourth replicas.
Thus, the only terms that survive the application of the first-- and second--subsystem component of $A_\tau$ read
\be
\newket{\omega_1\\ \omega_2\\ \omega_1\\ \omega_2}\ ,\ \newket{\omega_1\\ \omega_2\\ \omega_2\\ \omega_1}\ ,\
\newket{\omega_2\\ \omega_1\\ \omega_1\\ \omega_2}\ \mbox{and}\ \newket{\omega_2\\ \omega_1\\ \omega_2\\ \omega_1}\ .
\label{eq:w_states}
\ee
However, all these terms are completely symmetric in the third subsystem components,
and, therefore, are mapped onto the zero-vector by $P_-^{(13)}\otimes P_-^{(24)}$ on ${\cal H}_3^{\otimes 4}$.
That is, all-together, we have verified that
\be
A_\tau\ket{W}^{\otimes 4}=0\ ,
\label{eq:AW}
\ee
what, together with Eq.~\eqref{eq:AGHZ}, veryfies that $GHZ$ states cannot be generated through local manipulation of $W$ states.
Since all reduced density matrices of both the $GHZ$ and the $W$ state have full rank,
this, in turn, implies the in-equivalence of these two states.

This reasoning can directly be generalized also to systems with more than three constituents,
where $\ket{GHZ^n}=(\ket{1}^{\otimes n}+\ket{2}^{\otimes n})/\sqrt{2}$ and $\ket{W^n}=\sum^n_i \ket{w_i}/\sqrt{n}$. Here, $\ket{w_i}$ is the product vector with the i-th subsystem in state $\ket{2}$ and all others in $\ket{1}$. These states are the natural generalization of the GHZ and W states to an $n$-qubit system.
Replacing $A_\tau$ by $A_{\tau} \otimes \mathbbm1$, with $\mathbbm1$ the identity in the quadruplicated Hilbert space of $n-3$ subsystems, yields the analogue of Eq.~\eqref{eq:AGHZ} and Eq.~\eqref{eq:AW} for systems with arbitrarily many subsystems.

%------------------------------------------------------------------------------------------------------------------------------

\subsection{Beyond qubits}

Having identified the distinctness of $GHZ$ and $W$ states, we can now leave the realm of two-level systems
and identify in a similar fashion inequivalent states in higher-dimensional systems.
Natural generalizations of these states to three-level systems with full rank of all reduced density matrices are
\be
\ket{GHZ_3}=\frac{1}{\sqrt{3}}\bigl(\ket{111}+\ket{222}+\ket{333}\bigr)\ ,
\ee
and
\be
\ket{W_3}=\frac{1}{\sqrt{6}}\bigl(\ket{211}+\ket{121}+\ket{112}+\ket{322}+\ket{232}+\ket{223}\bigr)\ .
\ee
The operator $A_\tau$ as defined in Eq.~\eqref{eq:Atau} maps the quadruplication of both of these states on something finite,
{\it i.e.} it is {\em not} suited to show the inequivalence of these two states.
% We have to find on operator of stronger selectiveness under whose application replicas of exactly one of the two states vanish.
To find a suitable operator, we can resort to the projector ${\cal A}_3$ onto the completely anti-symmetric component of a triplicate Hilbert space defined above in Eq. \eqref{eq:defantisymmetrizer3}
in the context of bipartite systems.
In terms of this projector -- that we will simply denote ${\cal A}$ in the remainder -- we can define the generalization of $A_\tau$ to
\be
A_{3\tau}=\underbrace{{\cal A}^{(123)}\otimes {\cal A}^{(456)}}_{{\cal H}_1^{\otimes 6}}\otimes\underbrace{{\cal A}^{(125)}\otimes {\cal A}^{(346)}}_{{\cal H}_2^{\otimes 6}}\otimes\underbrace{{\cal A}^{(134)}\otimes {\cal A}^{(256)}}_{{\cal H}_3^{\otimes 6}}\ ,
\label{eq:tangle_3level}
\ee
where the indices $ijk$ of ${\cal A}^{(ijk)}$ denote the $i$-th, $j$-th and $k$-th replica of the respective single-particle Hilbert space
as also illustrated in Fig.~\ref{fig:sym_tangle}.
\begin{figure}
\caption{On six replicas of a tripartite Hilbert space, with three-dimensional subsystems, one can define the operator $A_{3\tau}$ in
Eq.~\eqref{eq:tangle_3level}, to show the inequivalence of three-level $GHZ$ and $W$ states.
It is based on projectors ${\cal A}$ onto the completely anti-symmetric component of triples of local Hilbert spaces.
Each of the triples that enter Eq.~\eqref{eq:tangle_3level} are identified by connecting black lines.}
\label{fig:sym_tangle}
\end{figure}

With this operator, we can explicitly verify that
\be
A_{3\tau}\ket{W_3}^{\otimes 6}=0\ ,\ \mbox{but}\ ,\ A_{3\tau}\ket{GHZ_3}^{\otimes 6}=\frac{1}{4 \cdot 3^5}\ket{\xi}\ ,
\ee
where $\ket{\xi}$ is the unique eigenvector associated with the unit eigenvalue of $A_{3\tau}$.

As discussed in the context of Eq.~\eqref{eq:signature}, only three-replica state-vectors with pairwise different components survive the application of ${\cal A}$.
Consequently, if there was such a term in the six-fold $W_3$-state, $\ket{W_3}^{\otimes 6}$, it necessarily would need to contain each of the terms $\ket{1}$, $\ket{2}$ and $\ket{3}$ exactly twice per subsystem. The only terms in $\ket{W_3}$ that contain the term $\ket{3}$ are $\ket{322}$, $\ket{232}$ and $\ket{223}$.
If a term in $\ket{W_3}^{\otimes 6}$ is to contain $\ket{3}$ exactly twice per subsystem, {\it i.e.} twice in each column in Fig.~\ref{fig:sym_tangle}, it must solely consist of a six-fold tensor product of the terms $\ket{322}$, $\ket{232}$ and $\ket{223}$, each entering twice. Therefore any such term in $\ket{W_3}^{\otimes 6}$ does not contain the term $\ket{1}$,  that is, contains no contribution of the completely anti-symmetric three-level state, and therefore is necessarily mapped to the zero-vector by $A_{3\tau}$.

The sixfold $\ket{GHZ_3}$ reads
$\ket{GHZ_3}^{\otimes 6}=\sum_{i,j,k,p,q,l=1}^3\ket{iii}\otimes\ket{jjj}\otimes\ket{kkk}\otimes\ket{lll}\otimes\ket{ppp}\otimes\ket{qqq}/27$,
but as we discuss in the following, only a few of the overall $3^6$ terms yield a finite contribution after the application of $A_{3\tau}$.
Due to ${\cal A}^{(123)}$ acting on the first subsystem the first three replicas need to contain three pairwise different components, that we label $i$, $j$ and $k$.
Because of ${\cal A}^{(125)}$ acting on the second subsystem, the fifth replica needs to be in the state $\ket{k}$ since the first replicas are in the states $\ket{i}$ and $\ket{j}$.
Similarly, the fourth replica needs to be in $\ket{j}$, due to the action of ${\cal A}^{(134)}$ on the third subsystem.
Finally, the sixth replica needs to be in state $\ket{i}$, since ${\cal A}^{(456)}$ acts on the first subsystems.
Consequently, all terms in $\ket{GHZ_3}^{\otimes 6}$ that are not mapped on the zero vector by $A_{3\tau}$ are of the form
\be
\newket{i\\j\\k\\j\\k\\i}\otimes\newket{i\\j\\k\\j\\k\\i}\otimes\newket{i\\j\\k\\j\\k\\i}\ ,
\label{GHZterms}
\ee
What remains to be shown is that the six ($3!$) different terms with pairwise different $i$, $j$ and $k$ don't add up destructively.
For this purpose, we have to inspect the prefactors as given in Eq.~\eqref{eq:signature}, which one obtains through the application of the six different terms ${\cal A}$ in $A_{3\tau}$.
In the first subsystem, there is ${\cal A}$ acting on the first three replicas and on the last three replicas.
In these two terms, we have
\be
\ket{\kappa_1}=\newket{i\\j\\k}\ ,\mbox{and}\ \ket{\kappa_2}=\newket{j\\k\\i}\ .
\ee
In the second subsystem, ${\cal A}$ acts on replicas $1$, $2$ and $5$ and on replicas $3$, $4$ and $6$.
In these two terms, we have
\be
\ket{\kappa_3}=\newket{i\\j\\k}\ ,\mbox{and}\ \ket{\kappa_4}=\newket{k\\j\\i}\ .
\ee
In the third subsystem, ${\cal A}$ acts on replicas $1$, $3$ and $4$ and on replicas $2$, $5$ and $6$.
In these two terms, we have
\be
\ket{\kappa_5}=\newket{i\\k\\j}\ ,\mbox{and}\ \ket{\kappa_6}=\newket{j\\k\\i}\ .
\ee
The states $\ket{\kappa_1}$, $\ket{\kappa_2}$, $\ket{\kappa_3}$ and $\ket{\kappa_6}$ have the same signature,
whereas the states $\ket{\kappa_4}$ and  $\ket{\kappa_5}$ have the opposite signature.
But, since both signatures appear an even number of times, the overall prefactor is always positive,
so that the different terms of Eq.~\eqref{GHZterms} add up constructively after the application of $A_{3\tau}$.
With this, we have shown the inequivalence of the $\ket{W_3}$ and the $\ket{GHZ_3}$ state.

These two states, however, certainly do not define the only classes of entanglement in tripartite three-level systems.
For example, there is the Aharonov state
$\ket{\chi_3}=1/\sqrt{6}((\ket{123}+\ket{231}+\ket{312})-(\ket{132}+\ket{213}+\ket{321}))$ \cite{PhysRevLett.87.217901},
that also survives $A_{3\tau}$
\be
A_{3\tau}\ket{\chi}^{\otimes 6}=\frac{1}{2 \cdot 6^3}\ket{\xi}\ .
\ee
That is, whereas this shows the inequivalence of $W_3$ and Aharonov state,
we need to invoke another symmetry operation to show the inequivalence of $GHZ_3$ and Aharonov state.
For this purpose, we use the 
the projector
\be
{\cal S}=\frac{1}{3!}\sum_{i=1}^{3!}\Pi_{i}
\ee
onto the completely symmetric (bosonic) part of a triplicate single particle Hilbert space.
Similarly to Eq.~\eqref{eq:signature} above, ${\cal S}$ maps a tri-replica state-vector with components $\ket{1}$, $\ket{2}$ and $\ket{3}$
on a completely symmetric state
\be
\ket{{\cal S}}=\frac{1}{\sqrt{6}}\left(\left|\ba{c}1\\2\\3\\\ea\right\rangle+\left|\ba{c}1\\3\\2\\\ea\right\rangle+\left|\ba{c}2\\1\\3\\\ea\right\rangle+\left|\ba{c}2\\3\\1\\\ea\right\rangle+\left|\ba{c}3\\1\\2\\\ea\right\rangle+\left|\ba{c}3\\2\\1\\\ea\right\rangle\right)\ .
\ee
With the help of the operators ${\cal A}$ and ${\cal S}$, each acting on three replicas of one subsystem, we can now define the operator
\be
\underbrace{\cal A}_{{\cal H}_1^{\otimes 3}}\otimes\underbrace{\cal A}_{{\cal H}_2^{\otimes 3}}\otimes\underbrace{\cal S}_{{\cal H}_3^{\otimes 3}}\ ,
\label{eq:AAS}
\ee
that permits the discrimination of $GHZ_3$ and Aharonov state.
The triplication of the $GHZ_3$ state reads
$\ket{GHZ_3}^{\otimes 3}=\sum_{i,j,k=1}^3\ket{iii}\otimes\ket{jjj}\otimes\ket{kkk}/\sqrt{3}^3$.
All terms in this sum with $i$, $j$ and $k$ not mutually different are mapped on the zero-vector by ${\cal A}$,
since there is no completely anti-symmetric three-particle state in less than three dimensions.
The only terms that are not mapped to zero are of the form
\be
\newket{i\\j\\k}\otimes\newket{i\\j\\k}\otimes\newket{i\\j\\k}\ ,
\label{eq:tripleGHZ}
\ee
with $i,j,k$ pairwise different.
The application of ${\cal A}\otimes{\cal A}\otimes{\cal S}$ on this terms yields
\be
{\cal A}\left|\ba{c}i\\j\\k\\\ea\right\rangle\otimes{\cal A}\left|\ba{c}i\\j\\k\\\ea\right\rangle\otimes{\cal S}\left|\ba{c}i\\j\\k\\\ea\right\rangle=\frac{1}{\sqrt{6}^3}\mbox{sgn}^2(ijk)\ \ket{{\cal A}}\otimes\ket{{\cal A}}\otimes\ket{{\cal S}}.
\ee
Since all terms enter with a positive prefactor, {\it i.e.} add constructively, the application of ${\cal A}\otimes{\cal A}\otimes{\cal S}$ on the triplicated $GHZ$ state results in something finite:
\be
{\cal A}\otimes{\cal A}\otimes{\cal S}\ket{GHZ_3}^{\otimes 3}=\frac{1}{9\sqrt{2}}\ket{{\cal A}}\otimes\ket{{\cal A}}\otimes\ket{{\cal S}}\ .
\label{eq:AASGHZ3}
\ee

This is different for the Aharonov state:
let us review the terms of $\ket{\chi}^{\otimes 3}$ in which the terms associated with the three replicas of the first and second subsystems simultanuously have pairwise different terms, since only those survive the projection ${\cal A} \otimes {\cal A} $ onto the completely anti-symmetric parts.
Labelling the replica components of the first subsystems $\ket{i}$, $\ket{j}$ and $\ket{k}$ (pairwise different),
we obtain the two different terms:
\be
\ket{\chi_a}=\newket{i\\j\\k}\otimes\newket{j\\k\\i}\otimes\newket{k\\i\\j}\ ,
\label{eq:sigpos}
\ee
and
\be
\ket{\chi_b}=\newket{j\\k\\i}\otimes\newket{i\\j\\k}\otimes\newket{k\\i\\j}\ .
\label{eq:signeg}
\ee
All factors associated with a single replica component in $\ket{\chi_a}$ have the same signature,
and the same also holds for all single replica components in $\ket{\chi_b}$.
However, the signature of the $\ket{\chi_a}$--components is different from the signature of the $\ket{\chi_b}$--components,
and since the number of replicas is odd, $\ket{\chi_a}$ enters $\ket{\chi}^{\otimes 3}$ with a different sign than $\ket{\chi_b}$.
Similarly, the first subsystem-components of $\ket{\chi_a}$ have the same signature as the second subsystem components,
and the same holds true for $\ket{\chi_b}$.
That is, independently of what this signature is,
the contribution from the application of ${\cal A}$ on the first subsystem is canceled by the contribution of ${\cal A}$ on the second subsystem, {\it i.e.}
\be
{\cal A}\newket{i\\j\\k}\otimes{\cal A}\newket{j\\k\\i}={\cal A}\newket{i\\j\\k}\otimes{\cal A}\newket{k\\i\\j}=\frac{1}{6}\ket{\cal A}\otimes\ket{\cal A}\ .
\ee
Since, however, $\ket{\chi_a}$ and $\ket{\chi_b}$ enter $\ket{\chi}^{\otimes 3}$ with a different sign,
and the third subsystem components are always mapped on the symmetric state $\ket{{\cal S}}$,
these terms cancel after the application of ${\cal A}\otimes{\cal A}\otimes{\cal S}$,
so that
\be
{\cal A}\otimes{\cal A}\otimes{\cal S}\ket{\chi}^{\otimes 3}= 0\ .
\label{eq:chidies}
\ee

Quite surprisingly, we can also see that a biseparable state $\ket{\Psi_{BS}}=\sum_{i=1}^{3}\sqrt{\lambda_i}\ket{ii\varphi}$, with a bipartite entangled component of full Schmidt rank,
can not be obtained from the Aharonov state,
since the application of ${\cal A}\otimes{\cal A}\otimes{\cal S}$ yields
\be
{\cal A}\otimes{\cal A}\otimes{\cal S}\ket{\Psi_{BS}}^{\otimes 3}=(\lambda_1\lambda_2\lambda_3)^{3/2}\frac{1}{6\sqrt{6}}\ket{\cal A}\otimes\ket{\cal A}\otimes\newket{\varphi\\\varphi\\\varphi}\ ,
\ee
{\it i.e.} something finite.
If $\lambda_3$ vanishes, {\it i.e.} if the state exhibits only two-qubit entanglement, this component vanishes, and then, indeed, the biseparable state can be obtained from the Aharonov state.
That is, whereas in the case of qubits tripartite entanglement is in general superior to bipartite entanglement,
{\it i.e.} all bipartite qubit entangled states can be obtained from both W and GHZ entangled states, this is no longer true for higher dimensional systems.
%----------------------------------------------------------------------------------------------------------------------------------------------------------------------------------------------------------------------------------

Finally, we would like to illustrate that the criterion Eq.~\eqref{eq:cond} is capable of proving the inequivalence of two higher-dimensional multipartite states even when it is solely due to distinct phases in their respective product decompositions. Consider the Aharonov state $\ket{\chi_3}$ on the one side and its symmetric counterpart $\ket{\chi^+_3}=1/\sqrt{6}((\ket{123}+\ket{231}+\ket{312})+(\ket{132}+\ket{213}+\ket{321}))$ on the other. Both states have equivalent Schmidt decompositions with respect to any bipartite splitting: both, the Schmidt coefficients and the entanglement properties of the Schmidt bases coincide.\\
As discussed above between Eqs.~\eqref{eq:AAS} and \eqref{eq:chidies}, $\ket{\chi_3}^{\otimes 3}$ vanishes under the application of ${\cal A}\otimes{\cal A}\otimes{\cal S}$,
due to the destructive interference of the terms $\ket{\chi_a}$ and $\ket{\chi_b}$ defined in Eqs.~\eqref{eq:sigpos} and \eqref{eq:chidies}.
In the case of $\ket{\chi^+_3}$, however, these terms interfere constructively, so that
\be
{\cal A}\otimes{\cal A}\otimes{\cal S}\ket{\chi^+_3}^{\otimes 3}=\frac{1}{36}\ket{{\cal A}}\otimes\ket{{\cal A}}\otimes\ket{{\cal S}}\ ,
\ee
what proves the inequivalence of $\ket{\chi_3}$ and $\ket{\chi^+_3}$.

\subsection{N-Partite systems}

Having identified symmetries that permit to distinguish different states of tripartite three-level systems,
one can also generalize such classifications to systems of general particle number.
Here, we would like to demonstrate this for the case of the $GHZ$ and the Aharonov states for $N$-partite $N$-level systems for general odd $N$.
%We will now expand the above results to generalizations of the $GHZ$ and the Aharonov states, to n-partite systems, each subsystem of dimension N, for all odd numbers N,
The generalizations of the states read
\be
\ket{GHZ_N}=\frac{1}{\sqrt{N}}\sum_i^N \ket{i}^{\otimes N}\ ,
\label{eq:GHZN}
\ee
and
\be
\ket{\chi_N}=\frac{1}{\sqrt{N!}}\sum_{i_1i_2...i_N}^N \epsilon_{i_1i_2...i_N}\ket{i_1i_2...i_N}\ ,
\label{eq:AharonovN}
\ee
where $\epsilon_{i_1i_2...i_N}$ is the completely anti-symmetric tensor.

The operator that achieves the desired distinction % between these two states reads
\be
P_N=\underbrace{\cal A_N}_{{\cal H}_1^{\otimes N}}\otimes\underbrace{\cal A_N}_{{\cal H}_2^{\otimes N}}\otimes\underbrace{\cal S_N}_{{\cal H}_3^{\otimes N}}\otimes...\otimes\underbrace{\cal S_N}_{{\cal H}_N^{\otimes N}}\ ,
\label{eq:AAmanyS}
\ee
with $\cal A_N$ as defined in Eq.~\eqref{eq:AN} and $\cal S_N$ the projector onto the completely symmetric subspace of ${\cal H}^{\otimes N}$.
Analogously to \eqref{eq:AASGHZ3} one obtains
\be
P_N\ket{GHZ_N}^{\otimes N}=\frac{1}{\sqrt{N}\sqrt{N!}^{N-2}}\ket{{\cal  A_N  A_N  S_N...  S_N}}\ ,
\ee
where $\ket{{\cal A_N}}$ and $\ket{{\cal S_N}}$ are the completely anti-symmetric and completely symmetric state of an $N$-partite $N$-level system respectively.

Also the verification of $P_N\ket{\chi_N}^{\otimes N}=0$ is very analogous to the three-body case discussed above.
Similarly to Eqs.~\eqref{eq:sigpos} and \eqref{eq:signeg}, the only state-vectors that are not immediately mapped on the zero-vector are of the form
\be
\ket{\chi_{aN}}=\ket{\vec x}\otimes\ket{\vec y}\otimes\ket{\Phi_{N-2}}\ ,
\ee
where $\ket{\vec x}$ is a shorthand notation for an $N$-replica state and all elements of the vectors $\vec x$ and $\vec y$ need to be pariwise different, as well as $\vec x\neq \vec y$.
$\ket{\Phi_{N-2}}$ denotes the state of the $N$ replica of the last $N-2$ subsystems.
To each such state, there is the corresponding state
\be
\ket{\chi_{bN}}=\ket{\vec y}\otimes\ket{\vec x}\otimes\ket{\Phi_{N-2}}\ .
\ee
both states share the same $\ket{\Phi_{N-2}}$ and since ${\cal A}\ket{\vec x}=sgn(\vec x)N!^{-1/2}\ket{{\cal A}}$, it follows that ${\cal A}\otimes {\cal A}\ket{x}\otimes\ket{\vec y}={\cal A}\otimes {\cal A}\ket{y}\otimes\ket{\vec x}$. Consequently, one obtains
\be
P_N\ket{\chi_{aN}}=P_N\ket{\chi_{bN}}\ .
\ee
Yet, for odd N, the terms $\ket{\chi_{aN}}$ and \ket{\chi_{bN}} appear with different sign in $\ket{\chi_N}$; the two terms $P_N\ket{\chi_{aN}}$ and $P_N\ket{\chi_{bN}}$ cancel each other, what implies that $P_N\ket{\chi_N}^{\otimes N}=0$.

\section{Outlook}

The above exemplary cases of inequivalent multipartite states, as well as the reproduction of prior classification
with one and the same recipe, raise confidence that Eq.~\eqref{eq:cond}, as originally established in \cite{PhysRevA.77.052304},
will eventually provide a systematic understanding of the general structure of multi-partite entanglement.
Whereas we do not rigorously know whether Eq.~\eqref{eq:cond} is sufficient to provide an exhaustive characterization of all inequivalent classes of multi-partite entangled states, the following arguments suggest that this indeed might be the case. On the one hand, the rapid growth of the number of different permutations with the number $M$ of a state's multiplicity, which comes in hand with a growing variety of independent terms in Eq.~\eqref{eq:A}, results in an enhanced freedom to build suitable operators $A$. 
 
We focussed here exclusively on the classification of multi-partite entanglement, but our presently described framework bears the potential for much broader applications.
Consider for example the question of filtering \cite{Gisin1996151} or entanglement distillation \cite{PhysRevLett.76.722},
where local Kraus operators $F_i$ that map a mixed state $\Sigma$ to a given pure state $\ket{\Psi}$,
{\it i.e.} $F_1\otimes\hdots\otimes F_N\Sigma F_1^\dagger\otimes\hdots\otimes F_N^\dagger=\ket{\Psi}\bra{\Psi}$,
are sought.
Here, $\Sigma$ can be the state of an $N$-partite quantum system, in the case of filtering, or, it can be a multiple tensor product of the to-be-distilled state $\varrho$,
{\it i.e.} $\Sigma=\varrho^{\otimes q}$, where $q$ subsystems define a single logical subsystem as in regular distillation.
In this context, a relation $A\Sigma^{\otimes M} A^\dagger=0$, but $A\ket{\Psi}^{\otimes M}\neq 0$ proves the non-existence of a filtering- or distillation-protocoll.
% {\bf This can, in principle, be experimentally verified if A is a projector, which is indeed the case throughout this manuscript, since it defines a measurement description, see, e.g. \cite{PhysRevLett.97.050501}.}

Very naturally, the present framework also facilitates the construction of entanglement measures that target very specific entanglement properties:
considering expectation values of the symmetry operations $A$ defined here, one arrives at scalar functions --
just like the tangle \cite{PhysRevA.61.052306}, that can be defined in terms of Eq. \eqref{eq:Atau},
or, generalized concurrences \cite{gour:012318}, that can be defined in terms of ${\cal A}$.
Once verified the equality $A\ket{\Phi}^{\otimes M}=0$,
one assures that the function $\bra{\Psi}^{\otimes M}A\ket{\Psi}^{\otimes M}$ is completely insensitive to entanglement of the class of $\ket{\Phi}$,
what enables the construction of entanglement measures that access specific entanglement properties,
% On the other hand, any \emph{quantitative} entanglement property of a quantum state, {\it i.e.} an entanglement measure, can be expressed in form of expectation values of these operators $A$ as well \cite{PhysRevA.58.1833} {\bf: passt diese Referenz hier},
and those measures are directly accessible in experiments on multiple identically prepared quantum systems \cite{aolita:050501}.

\vfill

\begin{acknowledgements}
We acknowledge stimulating discussions with Felix Eckert, Gregor Schaumann and Tsubasa Ichikawa.
F.M. acknowledges financial support by Deutsche Forschungsgemeinschaft (DFG) under the grant MI 1345/2-1.
\end{acknowledgements}
\bibliography{referenzen}

\end{document}